\documentclass[aps,prd,preprint,floatfix,nofootinbib,superscriptaddress]{revtex4}

\usepackage{graphicx,subfigure,epsfig,epsf}

\newcommand{\s}{\sigma}
\newcommand{\g}{\gamma}
\newcommand{\om}{\omega}

\newcommand{\beq}{\begin{equation}}
\newcommand{\eeq}{\end{equation}}
\newcommand{\bea}{\begin{eqnarray}}
\newcommand{\eea}{\end{eqnarray}}

\def\Re{{\cal R \mskip-4mu \lower.1ex \hbox{\it e}\,}}
\def\Im{{\cal I \mskip-5mu \lower.1ex \hbox{\it m}\,}}
\def\ie{{\it i.e.}}
\def\eg{{\it e.g.}}

\def\etal{{\it et al.}}

\def\tev{\,{\ifmmode\mathrm {TeV}\else TeV\fi}}
\def\gev{\,{\ifmmode\mathrm {GeV}\else GeV\fi}}
\def\mev{\,{\ifmmode\mathrm {MeV}\else MeV\fi}}
\def\to{\rightarrow}

\begin{document}

% Reference Macros:  Enter parameters in order Vol, Page, Year
%\def\issue(#1,#2,#3){{\bf #1}, #2 (#3)} % AIP format

\def\issue(#1,#2,#3){#1 (#3) #2} % PLB format
\def\APP(#1,#2,#3){Acta Phys.\ Polon.\ \issue(#1,#2,#3)}
\def\ARNPS(#1,#2,#3){Ann.\ Rev.\ Nucl.\ Part.\ Sci.\ \issue(#1,#2,#3)}
\def\CPC(#1,#2,#3){comp.\ Phys.\ comm.\ \issue(#1,#2,#3)}
\def\CIP(#1,#2,#3){comput.\ Phys.\ \issue(#1,#2,#3)}
\def\EPJC(#1,#2,#3){Eur.\ Phys.\ J.\ C\ \issue(#1,#2,#3)}
\def\EPJD(#1,#2,#3){Eur.\ Phys.\ J. Direct\ C\ \issue(#1,#2,#3)}
\def\IEEETNS(#1,#2,#3){IEEE Trans.\ Nucl.\ Sci.\ \issue(#1,#2,#3)}
\def\IJMP(#1,#2,#3){Int.\ J.\ Mod.\ Phys. \issue(#1,#2,#3)}
\def\JHEP(#1,#2,#3){J.\ High Energy Physics \issue(#1,#2,#3)}
\def\JPG(#1,#2,#3){J.\ Phys.\ G \issue(#1,#2,#3)}
\def\MPL(#1,#2,#3){Mod.\ Phys.\ Lett.\ \issue(#1,#2,#3)}
\def\NP(#1,#2,#3){Nucl.\ Phys.\ \issue(#1,#2,#3)}
\def\NIM(#1,#2,#3){Nucl.\ Instrum.\ Meth.\ \issue(#1,#2,#3)}
\def\PL(#1,#2,#3){Phys.\ Lett.\ \issue(#1,#2,#3)}
\def\PRD(#1,#2,#3){Phys.\ Rev.\ D \issue(#1,#2,#3)}
\def\PRL(#1,#2,#3){Phys.\ Rev.\ Lett.\ \issue(#1,#2,#3)}
\def\PTP(#1,#2,#3){Progs.\ Theo.\ Phys. \ \issue(#1,#2,#3)}
\def\RMP(#1,#2,#3){Rev.\ Mod.\ Phys.\ \issue(#1,#2,#3)}
\def\SJNP(#1,#2,#3){Sov.\ J. Nucl.\ Phys.\ \issue(#1,#2,#3)}

\begin{flushright}
BITSGoa-2008/01/001\\
\end{flushright}

%\input titlepage

% You should use BibTeX and revtex.bst for references
\bibliographystyle{revtex}

\title{Plasmon Annihilation into Kaluza-Klein Graviton: New Astrophysical Constraints on Large Extra Dimensions} 
% Optional argument for running titles on pages
%\title[]{}


% \affiliation command. The \affiliation command should follow th% other information

%\email[]{email}
%%\homepage[]{Your web page}
%%\thanks{}
%%\altaffiliation{}
\author{Prasanta Kumar Das}
\email[]{pdas@bits-goa.ac.in}
\affiliation{Birla Institute of Technology and Science-Pilani, Goa Campus, NH-17B, Zuarinagar, Goa 403726, India.}
%\author{Ramesh Kaul}
%\affiliation{The Institute of Mathematical Sciences, C.I.T Campus, Taramani, Chennai 600113, India}
%\email[]{kaul@imsc.res.in}
\author{V~H~Satheeshkumar}
\email[]{vhsatheeshkumar@gmail.com}
\affiliation{Department of Physics, Sri Bhagawan Mahaveer Jain College of Engineering, Jain Global Campus, Kanakapura Road, Bangalore 562 112, India.}
\affiliation{School of Physics, University of Hyderabad, Central University P.O., Gachibowli, Hyderabad 500 046, India. }
\author{P. K. Suresh}
\email[]{pkssp@uohyd.ernet.in}
\affiliation{School of Physics, University of Hyderabad, Central University P.O., Gachibowli, Hyderabad 500 046, India. }
\date{\today}

\begin{abstract}
In large extra dimensional Kaluza-Klein (KK) scenario, where the usual Standard Model (SM) matter is confined to a $3+1$-dimensional hypersurface called the $3$-brane and gravity can propagate to the bulk ($D=4+d$, $d$ being the number of extra spatial dimensions), the light graviton KK modes can be produced inside the supernova core due to the usual nucleon-nucleon bremstrahlung, electron-positron and photon-photon annihilations. This photon inside the supernova becomes plasmon due to the plasma effect. In this paper, we study the energy-loss rate of SN 1987A due to the KK gravitons produced from the plasmon-plasmon annihilation. We find that the SN 1987A cooling rate leads to the  conservative bound $M_D > 22.9$ TeV and $1.38$ TeV for the case of two and three space-like extra dimensions. 
\end{abstract}

%\noindent PACS:~ 97.60.Bw; 98.80.-k; 11.25.Hf 

%\maketitle must follow title, authors, abstract and \pacs
\maketitle
%%%%%%%%%%%%%%%%%%%%%%%%%%%%%%%%%%%%%%%%%%%%%%%%%%%%%%%%%%%%%%%%%%%%%%%%%%
%%%%%%%%%%%%%%%%%%%%%%%%%%%%%%%%%%%%%%%%%%%%%%%%%%%%%%%%%%%%%%%%%%%%%%%%%%

% \cite{Das:2005das}, Ref.~\cite{Das:2005das}
%\label{Eq:lagr}, Eq.~(\ref{Eq:lagr}),
\section{Introduction}
Recently it has been noted that the scale of quantum gravity, the four dimensional Planck scale $M_{Pl}(\sim 10^{16}$ TeV), is just a conjecture without much experimental support and the only  experimentally  verified  scale of gauge interactions in four  dimensions lies  within  the TeV scale. Therefore, the assumptions that gravitation becomes  strong at  the TeV scale,  while  the standard gauge interactions remain  confined to the four dimensional spacetime,  does  not  conflict with the today's  experimental data  
%\cite{ADD}. 
These  ideas solve the hierarchy problem without relying on supersymmetry or technicolour and the observed weakness of gravity at long distances is due to the presence of $d$ new spatial dimensions large compared to the electroweak scale. This can be inferred from the relation between the Planck scales of the $D=4+d$ dimensional theory $M_{D}$ and the  four dimensional theory $M_{Pl}$, which, for the toroidal compactification, is given by
\bea \label{Eq:planckscale}
M_{Pl}^2 = (2 \pi R)^d M_{D}^{d+2},
\eea 
where $R$ is the size of the extra dimensions. Putting $M_{D} \sim 1$ TeV then yields
\beq
R \sim 10^{\frac{30}{d} - 17} \mbox{cm}.
\eeq 
For $d=1$, $R \sim 10^{13}$ cm, this case is obviously excluded since it would modify Newtonian gravitation at solar-system distances. For $d=2$, we get $R \sim 1$ mm, which is precisely the distance where our present experimental measurement of gravitational strength stops. Clearly, while the gravitational force has not been directly measured beneath a millimeter, the success of the SM up to $\sim 100$ GeV implies that the SM fields can not feel these extra large dimensions, that is they are confined to only  ``3-brane", in the higher dimensional spacetime called ``bulk''. Summarizing, in this framework the universe is $D=4+d$  dimensional with Planck scale near the weak scale, with $d \ge 2$ new sub-millimeter sized dimensions where gravity and perhaps other fields can freely propagate, whereas the SM particles are  localised on a 3-brane in this higher-dimensional spacetime.

This theory predicts a variety of novel signals which can be tested using table-top experiments, collider experiments, astrophysical or cosmological observations. It has been pointed out that one of the strongest bounds on this physics comes from SN 1987A~\cite{ADD2}. Various authors have done calculations to place such constraints on the extra dimensions \cite{Cullen:1999hc,BHKZ,Hannestad:2001jv,VHS-PKS}, which we briefly discuss here. The graviton emission from plasmon-plasmon (photon inside plasma of the supernovae (SN) becomes massive and is called as plasmon) annihilation might have deep impact on the supernovae cooling and can significantly alter the bounds on $M_D$. Here we have investigated this possibility. This would be similar to Farzan's treatment of the Majoron emission in the supernova cooling process as a source of the upper bound on neutrino-Majoron coupling~\cite{farzan} and Raffelt's treatment on axion emission in photon photon collision \cite{Raffelt1}. Several other mechanism for the SN 1987A cooling
comprising the New Physics(beyond the Standard Model Physics) are already available in the literatute. Recently Das \cite{das} and others (see \cite{das} for other works) have explored the unparticle physics as a possible cooling mechanism of the supernovae SN 1987A, in which an unparticle stuff can be produced in the nucleon-nucleon bremstrahlung, electron-positron and photon-photon annihilations and thus cools down the temparature of SN 1987A.  

\section{Supernova Explosion and Cooling}

Supernovae come in two main observational varieties: Type II are those whose optical spectra exhibit Hydrogen lines and have less sharp peaks at maxima (of 1 billion solar luminosities), whereas the optical spectra for the Type I supernovae does not have any Hydrogen lines and it exhibits sharp maxima \cite{VHS}. Physically, there are two fundamental types of supernovae, based on what mechanism powers them: the thermonuclear supernovae and the core-collapse ones. Only supernovae Ia are thermonuclear type and the rest are formed by core-collapse of a massive star. 
The core-collapse supernovae are the class of explosions which mark the evolutionary end of massive stars ($M \ge 8\,M_\odot$). 
 The kinetic energy of the explosion carries about 1\% of the liberated gravitational binding energy of about 
$3\times10^{53}~{\rm ergs}$ and the remaining 99\% going into neutrinos. This powerful and detectable neutrino burst is the main astro-particle interest of core-collapse supernovae.

In the case of SN 1987A, about $10^{53}~{\rm ergs}$ of gravitational binding energy was released in few seconds and the neutrino fluxes were measured by Kamiokande \cite{Kamio} and IMB \cite{IMB} collaborations. Numerical neutrino light curves can be compared with the SN 1987A data where the measured energies are found to be ``too low''.  For example,
the numerical simulation in \cite{Totani:1997vj} yields time-integrated values $\langle E_{\nu_e}\rangle\approx13~{\rm MeV}$, $\langle E_{\bar\nu_e}\rangle\approx16~{\rm MeV}$, and $\langle E_{\nu_x}\rangle\approx23~{\rm MeV}$.  On the other hand, the data imply $\langle E_{\bar\nu_e}\rangle=7.5~{\rm MeV}$ at Kamiokande and 11.1~MeV at IMB~\cite{Jegerlehner:1996kx}.  Even the 95\% confidence range for Kamiokande implies $\langle E_{\bar\nu_e}\rangle<12~{\rm MeV}$.  Flavor oscillations would increase the expected energies and
thus enhance the discrepancy~\cite{Jegerlehner:1996kx}.  It has remained unclear if these and other anomalies of the SN 1987A neutrino signal should be blamed on small-number statistics, or point to a serious problem with the SN models or the detectors, or is there a new physics happening in supernovae?

Since we have these measurements already at our disposal, now if we propose some novel channel through which the core of the supernova can lose energy, the luminosity in this channel should be low enough to preserve the agreement of neutrino observations with theory. That is,
${\cal L}_{new\, channel} \le 10^{53}\, ergs\, s^{-1}.$
This idea was earlier used to put the strongest experimental upper bounds on the axion mass \cite{axions}. Here, we will consider the gravitons which can carry the energy from the core of the supernovae and escape into the bulk of the larger dimensional space.
 The constraint on luminosity of this process can be converted into a bound on the 4+d dimensional Planck scale $M_D$. Any mechanism which leads to significant energy-loss from the supernovae core immediately after bounce will produce a very different neutrino-pulse shape, and so will destroy this agreement, which in the case of axion is explicitly shown by Burrows's \etal~\cite{BBT}.
%---as demonstrated explicitly in the axion case by Burrows, Brinkmann, and Turner 
Raffelt has proposed a simple analytic criterion based on detailed supernova simulations~\cite{Raffelt}: if any energy-loss mechanism has an emissivity greater than $10^{19}$ ergs g$^{-1}$ s$^{-1}$ then it will remove sufficient energy from the explosion to invalidate the current understanding of Type-II supernovae's neutrino signal. 
 Similar arguments can be applied to other particles. The hypothetical majorons are one case in point~\cite{Hannestad:2002ff}.
\section{Constraints on Extra Dimensions}

The most restrictive limits on $M_D$ come from SN~1987A energy-loss argument.  If large extra dimensions exist, the usual four dimensional graviton is complemented by a tower of Kaluza-Klein (KK) states, corresponding to new phase space in the bulk.   The KK gravitons interact with the strength of ordinary gravitons and thus are not trapped in the supernovae core.  During the first few seconds after collapse, the core contains neutrons, protons, electrons, neutrinos and thermal photons(plasmons). There are a number of processes in which higher-dimensional gravitons can be produced. For the conditions that pertain in the core at this time (temperatures $T \sim 30-70$ MeV, densities $\rho \sim (3-10) \times 10^{14}$ g cm$^{-3}$), the relevant processes are shown below
\begin{itemize}
\item  Graviton($\cal{G}$) production in Nucleon-Nucleon Brehmstrahlung: 
$N + N \to N + N + \cal{G}$
\item  Graviton production in photon fusion: 
$\gamma + \gamma \to \cal{G}$
\item  Graviton production in electron-positron annihilation process: 
$e^{-} e^{+} \to \cal{G}$
\end{itemize} 

In the supernovae, nucleon and photon abundances are comparable (actually nucleons are somewhat more abundant). In the following we present the bounds derived by various authors using nucleon-nucleon bremhmstrahlung and in the next section we give detailed calculation for photon-photon annihilation (including the plasma effect inside supernovae) to KK graviton process. We believe that 
although the dominant contribution will still follow from nucleon-nucleon bremsstrahlung, however, because of the large uncertainties involved in such a process calculation inside the hot plasma, the reliable bound will follow from plasmon + plasmon $\rightarrow$ KK graviton process. It is worthwhile to mention
here that in this work we have not considered the effect of plasmon width in the final continuum KK state production, which we believe if be taken into account will not substantiably change our bound on $M_D$.
We will not discuss the electron-positron annihilation to KK graviton as it does not give any significant bounds.

\subsection{Nucleon-Nucleon Brehmstrahlung}

This is the dominant process relevant for the SN~1987A where the temperature is comparable to $m_\pi$ and so the strong interaction between N's is unsuppressed. This process can be represented as
\beq
 N + N \to N + N + \cal{G},
\eeq
where $N$ can be a neutron or a proton and $\cal{G}$ is a higher-dimensional graviton. 

The main uncertainty comes from the lack of precise knowledge of temperatures in the core: values quoted in the
literature range from 30 MeV to 70 MeV.  For  $T=30$ MeV  and $\rho=3 \times 10^{14}$ g cm$^{-3}$, we list the results of various authors. \\
Cullen and Perelstein \cite{Cullen:1999hc}
\bea
M_D \, &\gtrsim& \, 50 \,\, \hbox{TeV},\hskip2cm d=2;\\
M_D \,&\gtrsim& \,\,\,\, 4 \,\, \hbox{TeV}, \hskip2cm d=3;\\
M_D \,&\gtrsim& \,\,\,\, 1 \,\, \hbox{TeV}, \hskip2cm d=4. 
\eea 
Barger, Han, Kao and Zhang \cite{BHKZ}
\bea
M_D \, &\gtrsim& \, 51 \,\, \hbox{TeV},\hskip2cm \,\,\, d=2;\\
M_D \,&\gtrsim& \,  3.6 \,\, \hbox{TeV},\hskip2cm \,    d=3.
\eea 
%%%%%%%%%%%%%%%%%%%%%%%%%%%%%%%%%%%%%%%%%%%%%%%%%%%%%%%%%%%%%%%%
%Hanhart, Phillips, Reddy, and Savage \cite{Hanhart:2001er}
%\bea
%R \,\lesssim \,7.1 \times \, 10^{-4}\hbox{mm},\hskip2cm\,\, n=2;\\
%R \, \lesssim \,8.5 \times 10^{-7} \,\, \hbox{mm}, \hskip2cm n=3.
%\eea 
%Hanhart, Phillips, Pons and Reddy \cite{Hanhart:2001fx}
%\bea
%M \, &\gtrsim& \, 31 \,\, \hbox{TeV},\hskip1.5cm R \,\lesssim \,6.6 \times \, 10^{-4}\hbox{mm},\hskip2cm\,\, n=2;\\
%M \,&\gtrsim& \,\,\,\, 2.75 \,\, \hbox{TeV}, \hskip1.5cm R \, \lesssim \,8.0 \times 10^{-7} \,\, \hbox{mm}, \hskip2cm n=3.
%\eea 
Hannestad and Raffelt \cite{Hannestad:2001jv}
\bea
M_D \, &\gtrsim& \, 84 \,\, \hbox{TeV},\hskip2cm d=2;\\
M_D \,&\gtrsim& \,\,\,\, 7 \,\, \hbox{TeV},\hskip2cm d=3.
\eea 
%Hannestad and Raffelt \cite{Hannestad:2003yd}
%\bea
%R \,\lesssim \,9.6 \times \, 10^{-4}\hbox{mm},\hskip2cm\,\, n=2;\\
%R \, \lesssim \,11.4 \times 10^{-7} \,\, \hbox{mm}, \hskip2cm n=3.
%\eea 
%%%%%%%%%%%%%%%%%%%%%%%%%%%%%%%%%%%%%%%%%%%%%%%%%%%%%%%%%%%%%%%%%%%%%%%%%%
\section{Methodology of calculation}
Each KK graviton state couples to the SM field with the $4$-dimensional gravitational strength according to \cite{Han:1999sg}
\beq
{\cal L}\ = -{\kappa\over2}\sum_{\vec n}
\int d^4 x\ h^{\mu\nu, {\vec n}} T_{\mu\nu}\ ,
\label{inter}
\eeq
where the summation is over all KK states labeled by the level $\vec n$. Here $\kappa = \sqrt{16 \pi G_N}$ and $G_N = 1/M_{Pl}^2$, the $4$-dimensional Newton's constant. $T_{\mu\nu}$ is the energy-momentum tensor of the SM  and $h^{\mu\nu, {\vec n}}$ the KK state. 

Since for large $R$ the KK gravitons are very light 
(because $m_{\vec n} \sim 1/R$), they may be copiously produced 
in high energy processes. For real emission of the KK gravitons 
from the collision of SM fields, the total cross-section can be written as
\beq
\sigma_{\rm tot}\ =\ \kappa^2\sum_{\vec n} \sigma({\vec n})\ ,
\eeq
where the dependence on the gravitational coupling is factored out.
Because the mass separation of adjacent KK states,  ${\cal O}(1/R)$,
is usually much smaller than typical energies in a physical process,
we can approximate the summation by an integration according to
\bea
\sum_{\vec n} \to \int \rho(m_{\vec n}^2) d(m_{\vec n}^2),
\eea
where the density of KK states $\rho(m_{\vec n}^2) = \frac{M_{Pl}^2}{M_D^{2+d}} \frac{1}{4^d \pi^{3d/2} \Gamma(d/2)} 
(m_{\vec n}^2)^{(d-2)/2}$. Here we have used the relation $M_{Pl}^2 = (2 \pi R)^d M_D^{2+d} $. 

 Now for a generic $2\to N$ body scattering, the scattering cross section is given by
\beq
\sigma= \frac{1}{Flux} \int \prod_f \frac{d^3 p_f}{(2 \pi)^3 2 E_f} (2 \pi)^4 \delta^4 \left( p_1+p_2 -\sum_f p_f\right) \overline{|{\cal{M}}_{fi}|^2}
\eeq 
%with
%\beq
%\upsilon_{rel}=\frac{\sqrt{(p_1\cdot p_2)^2-m_1^2 m_2^2}}{E_1 E_2},
%\eeq 
where $Flux= 4 E_1 E_2 \upsilon_{rel}$. Here $E_1$, $E_2$ are the energies of the initial particles 1 and 2 whose masses are $m_1$ and $m_2$, respectively and $\upsilon_{rel}$ is the relative velocity between them. 

For a general reaction of the kind $a+b \to c$, the above expression takes 
the form
\beq
\sigma = \frac{1}{Flux} \overline{|{\cal{M}}_{fi}|^2} 2 \pi \delta(S - m_c^2).
\label{cross}
\eeq

In the center-of-mass frame, we use the notation $\sqrt{S}$ for the total initial energy,
\bea
\sqrt{S}=E_1+E_2\label{cmenergy}\\
Flux = 4 E_1 E_2 \upsilon_{rel}=4 |\mathbf{p}| \sqrt{S},\label{relvel}
\eea 
where $|\mathbf{p}|=|\mathbf{p}_1|= |\mathbf{p}_2|= \frac{\lambda^{1/2}(S,m_1^2,m_2^2)}{2 \sqrt{S}}$ and $E_1$ and $E_2$ are the energies of the particles $a$ and $b$. The function $\lambda(x,y,z)(=x^2 + y^2 + z^2 - 2 x y -2 y z - 2 z x)$, is the standard  $K{\ddot{a}}$llen function.

Since we are concerned with the energy loss to gravitons escaping
into the extra dimensions, it is convenient and standard
\cite{Kolb,Raffelt} to define the quantities $\dot{\epsilon}_{a+b \to c}$
which are the rate at which energy is lost to gravitons via the
process $a + b \to c $ where $c$ has a decay width, 
per unit time per unit mass of
the stellar object. In terms of the cross-section $\sigma_{a+b \to c}$
the number densities  $n_{a,b}$ for a,b and the mass density
$\rho$, $\dot{\epsilon}$ is given by
\beq
\dot{\epsilon}_{a + b \to c.} = \frac{\langle n_a n_b
\sigma_{(a+b \to c)} v_{rel} E_{cm} \rangle}{\rho}
\label{emrate}
\eeq 
where the brackets indicate thermal averaging and $E_{cm}(=E_a + E_b)$ is the center-of-mass(c.o.m) energy of the two colliding particles $a$ and $b$.
Note that in the present case the final state KK graviton, although has smaller decay width but is  stable over the size of the neutron star because of it's large life time $\sim 10^9 (100 ~MeV/m)^3$ yr (See \cite{Han:1999sg}) and thus it can escape the supernovae while allowing it to cool.  
\section{Graviton production in plasmon fusion}

Photons are quite abundant in supernovae. Due to plasma effect inside the supernovae, photons becomes effectively massive. These massive photons(of mass $m_A$,say) are known as plasmons. Our interest is in the plasmon-plasmon annihilation to KK graviton \ie 
\beq
\gamma_P(k_1) + \gamma_P(k_2) \to KK(p).
\eeq 
%Fenman diagran for the reaction  is 
%\begin{center}
%\includegraphics[scale=1]{feynphoton.eps}
%\end{center}

The plasmon-plasmon-graviton ($G^n_{\mu\nu}(q) A^m_{\alpha}(k_1) A^{n-m}_{\beta}(k_2)$) vertex \cite{Han:1999sg} is given by
\bea
X_{\mu\nu\alpha\beta} &=& -\frac{i\kappa}{2}\biggl[(m_A^2 + k_1.k_2) C_{\mu\nu,\rho\sigma} +  D_{\mu\nu,\rho\sigma} (k_1, k_2) + \xi^{-1} E_{\mu\nu,\rho\sigma} (k_1, k_2) \biggr],
\eea
where the symbols $C_{\mu\nu,\rho\sigma}$,
$D_{\mu\nu,\rho\sigma}(k_1,k_2)$
, $E_{\mu\nu,\rho\sigma}(k_1,k_2)$ are defined as 
$$
C_{\mu\nu,\rho\sigma} = \eta_{\mu\rho}\eta_{\nu\sigma}
+\eta_{\mu\sigma}\eta_{\nu\rho}
-\eta_{\mu\nu}\eta_{\rho\sigma}\ ,
$$
$$
D_{\mu\nu,\rho\sigma} (k_1, k_2) =
\eta_{\mu\nu} k_{1\sigma}k_{2\rho}
- \biggl[\eta_{\mu\sigma} k_{1\nu} k_{2\rho}
  + \eta_{\mu\rho} k_{1\sigma} k_{2\nu}
  - \eta_{\rho\sigma} k_{1\mu} k_{2\nu}
  + (\mu\leftrightarrow\nu)\biggr]\ ,
$$
$$
E_{\mu\nu,\rho\sigma} (k_1, k_2) = \eta_{\mu\nu}(k_{1\rho}k_{1\sigma}
+k_{2\rho}k_{2\sigma}+k_{1\rho}k_{2\sigma})
-\left[\eta_{\nu\sigma}k_{1\mu}k_{1\rho}
+\eta_{\nu\rho}k_{2\mu}k_{2\sigma}
+(\mu\leftrightarrow\nu)\right].
$$
Here we work in the unitary gauge($\xi \to \infty$).
In the c.o.m frame, the momentum vectors for this reactions are
\bea
k_1^{\mu}&=& (E_1, 0, 0, k),\\
k_2^{\mu}&=& (E_2, 0, 0, -k),\\
p^{\mu} &=& (E_G, 0, 0, 0).
\eea 
It  often turns out to be more convenient to keep the polarizations explicitly. The polarization vectors \cite{Han:1999sg}
 of a massive graviton are
\bea
e_{\mu\nu}^{\pm 2}&=&2\epsilon_{\mu}^{\pm}\epsilon_{\nu}^{\pm}
\ , \nonumber\\
e_{\mu\nu}^{\pm 1}&=&\sqrt{2}\, (\epsilon_{\mu}^{\pm}\epsilon_{\nu}^{0}+
\epsilon_{\mu}^{0}\epsilon_{\nu}^{\pm})\ ,\nonumber\\
e_{\mu\nu}^{0}&=&\sqrt{\frac{2}{3}}\, (\epsilon_{\mu}^{+}
\epsilon_{\nu}^{-}+
\epsilon_{\mu}^{-}\epsilon_{\nu}^{+}-
2\epsilon_{\mu}^{0}\epsilon_{\nu}^{0})\ . \nonumber
\eea
Here $\epsilon_{\mu}^{\pm}$ and $\epsilon_{\mu}^{0}$ are the transverse and longitudinal polarization vectors of a massive gauge boson. For a massive vector
boson(\eg ~plasmon) with momentum $k^{\mu}=(E,0,0,k)$ and mass $m_A$,
\bea
 \epsilon^+_{\mu}(k)&=&\frac{1}{\sqrt{2}}(0,1,i,0)\ ,\\
 \epsilon^-_{\mu}(k)&=&\frac{1}{\sqrt{2}}(0,-1,i,0)\ ,\\
 \epsilon^0_{\mu}(k)&=&\frac{1}{m_A}(k,0,0,-E)\ .
\eea
The plasmon and graviton polarization vectors satisfy the following normalization and polarization sum conditions
\bea
e^{s\,\mu}e^{s'\, *}_{\mu} = 4\delta^{s s'},~~
\sum_{s=1}^3 e^s_{\mu}(k)e^{s\, *}_{\nu}(k) = -\eta_{\mu\nu} + \frac{k_\mu k_\nu}{m_A^2}\ ,\\
e^{s\,\mu\nu}e^{s'\, *}_{\mu\nu} = 4\delta^{s s'}\ ,~~
\sum_{s=1}^5 e^s_{\mu\nu}(p) e^{s\, *}_{\rho\sigma}(p) = B_{\mu\nu\, \rho\sigma}(p)\ ,
\eea
where $B_{\mu\nu\, \rho \sigma}(p)$ is given by
\bea
B_{\mu\nu\, \rho\sigma}(p) &=& 2\left(\eta_{\mu\rho}-\frac{p_\mu
p_\rho}{m_{\vec n}^2m_{\vec n}^2}\right) \left(\eta_{\nu\sigma}-\frac{p_\nu
p_\sigma}{m_{\vec n}^2}\right) \nonumber\\
&&+2\left(\eta_{\mu\sigma}-\frac{p_\mu
p_\sigma}{m_{\vec n}^2}\right)
\left(\eta_{\nu\rho}-{p_\nu p_\rho\over m_{\vec n}^2}\right)
\nonumber\\
&& - \frac{4}{3} \left(\eta_{\mu\nu}-\frac{p_\mu
p_\nu}{m_{\vec n}^2}\right) \left(\eta_{\rho\sigma}-\frac{p_\rho
p_\sigma}{m_{\vec n}^2}\right)\ .
\label{B}
\eea

The total squared amplitude, averaged over the initial three
polarizations(since massive plasmons have three state of polarizations) and summed over final states for the process 
$\gamma_P(k_1)+\gamma_P(k_2) \to G_{KK}(p)$ is
\bea \label{Eq:cross}
\overline {\left| {\cal M} \right|^2} = \left(\frac{1}{3}\right)^2 \sum_{s} \left| {\cal M}  \right|^2
 = {{\kappa^2}\over 72} \left(T_1^2 + T_2^2 + T_3^2 + T_4^2 + T_5^2\right), 
%\left( \frac{1}{m_A^4} \left(\frac{S^4}{12}-\frac{S^5}{12 m_{\vec n}^2} + \frac{S^6}{12 m_{\vec n}^4}\right) \right)
% \\
% + {{\kappa^2}\over 72}\left( \frac{1}{m_A^2} \left(\frac{5 S^4}{6 m_{\vec n}^2} - \frac{2 S^5}{3 m_{\vec n}^4}\right) 
\eea 
where $T_i^2 ~(i=1,..5)$ are given in appendix A. 
Substituting this in (\ref{cross}) and using (\ref{cmenergy}) and (\ref{relvel}), the total cross-section $\sigma_{T}$ for this process is obtained as
\bea \label{Eq:totcross}
\sigma_T = \sum_{\vec n} \sigma_{\g_P \g_P \rightarrow G_{kk}}(S,m_{\vec n})
 &=& \frac{1}{2 S}\int \rho(m_{\vec n}^2)~ \delta(S - m_{\vec n}^2)~ \overline {\left| {\cal M} \right|^2}~ d(m_{\vec n}^2) \nonumber \\
&=& \frac{1}{9}~ \frac{1}{4^d \pi^{z} \Gamma(d/2)}~  \left(\frac{S}{M_D^2}\right)^{d/2}~{\cal N}
\eea
where $z = -1 + \frac{3 d}{2}$ and  ${\cal N} = \frac{1}{M_D^2} \left(\frac{1}{12} \frac{S^2}{m_A^4} + \frac{1}{6} \frac{S}{m_A^2} + \frac{16}{3} \frac{m_A^2}{S} + 16 \frac{m_A^4}{S^2} + \frac{17}{3} \right) $. While deriving Eq.~\ref{Eq:totcross}, we have used $\rho(m_{\vec{n}})=\frac{R^d m_{\vec{n}}^{d-2}}{(4 \pi)^{d/2} \Gamma(d/2)}$ and the Planck scale relation Eq.~\ref{Eq:planckscale}. 

%Also, as mentioned earlier, $S$ is the center of mass energy, and $m_%{\vec n}$ the mass of the KK state at level $\vec n$.  
%%%%%START%%
%\beadf
%\sigma =\ {\pi\kappa^2 \sqrt{s}\over 16}
%\delta (m_{\vec n}-\sqrt{s})\ ,
%\eea 

The volume emissivity of a supernova  with a temperature $T$  through this process is obtained by thermal-averaging over the Bose-Einstein distribution. Hence, the energy loss rate  
($\dot{\epsilon}_{\g_P} = \frac{1}{\rho_{SN}}\dot{Q}_{\g_P}$) due to plasmon plasmon annihilation is given by (similar to that of the energy loss rate via 
$\gamma\gamma \to \nu\bar{\nu}$ \cite{ggnn})

\bea
\dot{\epsilon}_{\g_P} = \frac{1}{\rho_{SN}} \frac{1}{\pi^4}
\int_{\om_0}^{\infty} d \om_1 \frac{\om_1 (\om_1^2 - \om_0^2)^{1/2}}{e^{\om_1/T}-1}  \int_{\om_0}^{\infty}  d \om_2 \frac{\om_2 (\om_2^2 - \om_0^2)^{1/2}}{e^{\om_2/T}-1}
~{S (\om_1+\om_2)\over 2\om_1\om_2}~  \s_T,
\eea 
where $\s_T$ is given in Eq.~\ref{Eq:totcross}. 
Note that $N_{\g_P}= \frac{1}{\pi^2}
\int_{\om_0}^{\infty} d \om \frac{\om (\om^2 - \om_0^2)^{1/2}}{e^{\om/T}-1} $ is the number density of thermal photons, or rather of transverse plasmons. In the present case, we treat the plasmon to be transverse(with the dispersion relation given by  $\om^2 = \om_0^2 + |\bf{k}|^2$), since the contribution coming from the longitudinal plasmon is typically smaller \cite{Canuto,PR}. Also in above, $\om_0$ corresponds to plasma frequency in the supernovae core.

Finally introducing the dimensionless variables $x_i = \om_i/T$($i=0,1,2$) and taking $m_A$ (the transverse plasmon mass) to be equal to $\om_0$, we rewrite the above Equation as
\bea \label{Eq:lossrate}
\dot{\epsilon}_{\g_P} = \frac{1}{\rho_{SN}} 
\frac{T^{6+d}}{M_D^{2+d}\pi^4}
\int_{x_0}^{\infty} d x_1 \frac{x_1 (x_1^2 - x_0^2)^{1/2}}{e^{x_1/T}-1}  \int_{x_0}^{\infty}  d x_2 \frac{x_2 (x_2^2 - x_0^2)^{1/2}}{e^{x_2/T}-1}
~{(x_1+x_2)^{2+d} \over x_1 x_2}~  {\cal F},
\eea
where $${\cal F}=\frac{1}{18} \frac{1}{4^d \pi^z \Gamma(d/2)} 
\left[ 
\frac{T^4}{12 m_A^4} X_T^4 +  
\frac{T^2}{6 m_A^2} X_T^2  + 
\frac{16 m_A^2}{3 T^2} \frac{1}{X_T^2} + \\
\frac{16 m_A^4}{T^4} \frac{1}{X_T^4}  +
\frac{17}{3}
\right],~~X_T=x_1 + x_2.$$

\section{Numerical Analysis}
The SN 1987A energy loss due to KK graviton emission produced in massless photon-photon annihilation already put some bound on the effective scale of gravity $M_D$ for $d=2$ 
\noindent and $3$ (see \cite{BHKZ}). Here we study the modification of the above bound in a scenario where the plasma effect on photon is taken into account. In our analysis, the key working formula is the Eq.~\ref{Eq:lossrate} which describes the supernovae energy loss rate due to plasmon($\gamma_P$) + plasmon($ \gamma_P$) $\to$ KK graviton($G_{KK}$). Now for any kind of cooling mechanism which corresponds to an emissitivity $> 10^{19}~erg~g^{-1}~s^{-1}$ would invalidate our current understanding of Type-IIA supernovae's neutrino signal. 
So the consistency with the neutrino signal requires the energy loss rate 
$\le 10^{19}~ erg~ g^{-1} s^{-1}$. This gives rise the lower bound on $M_D$. In Fig. 1 we have shown the energy loss rate to KK gravitons as a function of  the scale $M_D$ for different number of extra dimensions $d$. The right and left curves respectively stands for $d=2$ and $3$. 
In this plot, the inputs taken are as follows: $\om_0 = m_A$(plasmon mass) $= 19$ MeV, the supernovae temperature  $T=30$ MeV and the supernovae core density 
$\rho \simeq 10^{15}$ $g~cm^{-3}$ \cite{Raffelt}. The horizontal line corresponds to the upper bound on the supernovae energy loss rate. The intersection of this curve with the other two gives rise the following lower bound on $M_D$: for $d=2$ we find $M_D > 22.9$ TeV, whereas for $d=3$, $M_D > 1.38$ TeV. The bound on $M_D$ as obtained here is somewhat stronger($d=2$) and weaker($d=3$) than that obtained in \cite{BHKZ} which are $15$ TeV and $1.6$ TeV, respectively for $d=2$ and $3$, where the relevant process of interest was photon-photon annihilation to KK gravitons. Also note that the bound on $M_D$ that we find from plasmon-plasmon annihilation to gravitons is somewhat weaker than the one obtained from the nucleon-nucleon brehmstrahlung which are  $51(3.6)$ TeV for $d=2(3)$, respectively. Finally, note that the present supernovae SN 1987A cooling analysis does not allow us to put any bound on $M_D$ for $d \ge 4$.

%%%%%%%%%%%%%%%%%%%%%%%%%%%%%%%%%%%%%%%%%%%%%%%%%%%%%%%%%%%%%%%%
\newpage
%\vspace*{-0.25in}
\begin{figure}
%\subfigure[Caption for picture one]{
\subfigure[]{
\label{PictureThreeLabel}
\hspace*{-0.87 in}
\begin{minipage}[b]{0.5\textwidth}
\centering
\includegraphics[width=\textwidth]{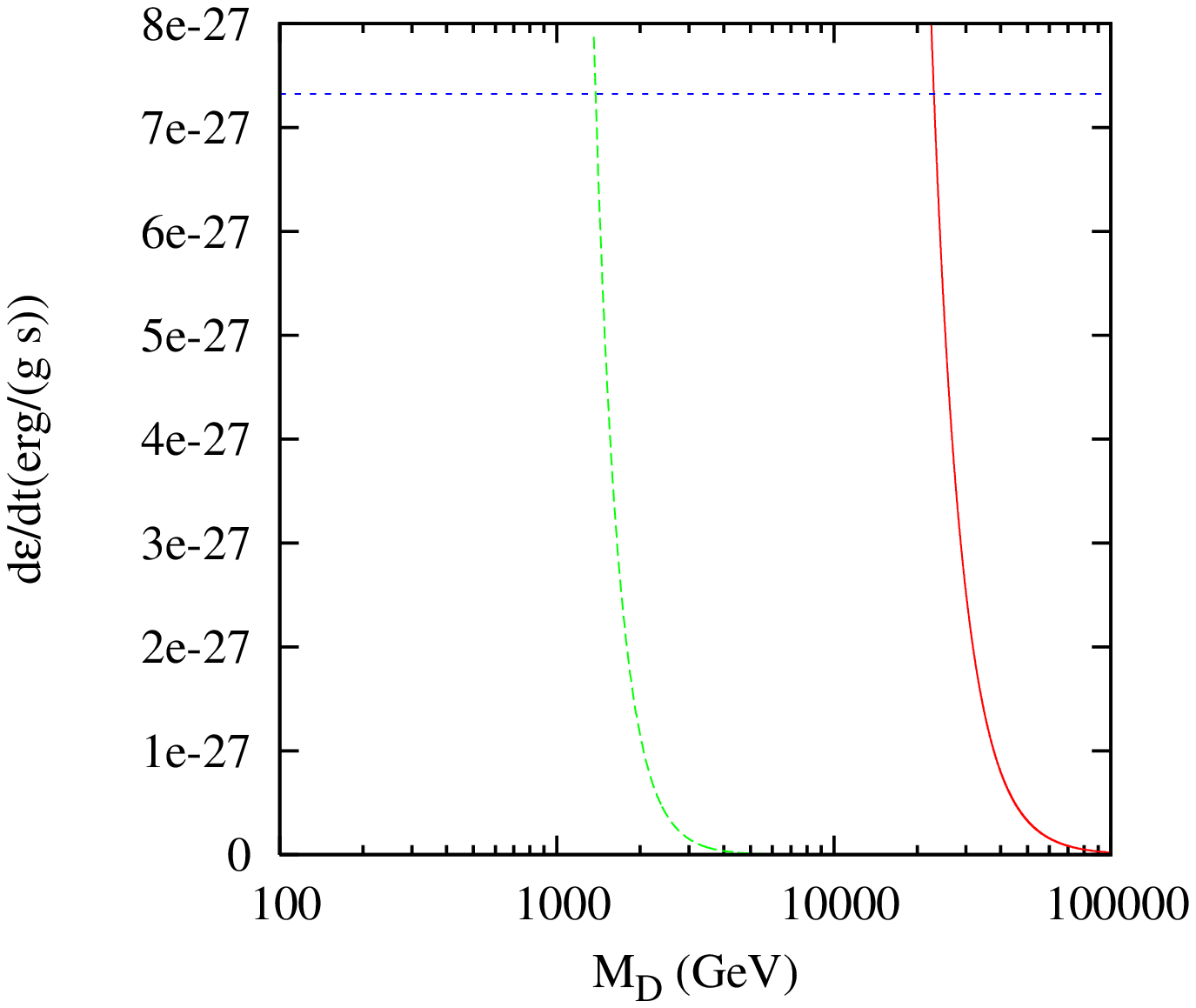}
\end{minipage}}
\end{figure}
\vspace*{-0.45in}
\noindent {\bf Fig. 1}.
{\it {The supernovae energy loss rate $d\epsilon/dt$ ($erg~ g^{-1} s^{-1}$) due to KK graviton emission produced in plasmon plasmon annihilation is shown as a function of $M_D$(GeV) in Fig. 1. For the right curve $d=2$, whereas for the left $d=3$. The upper horizontal curve corresponds to $d\epsilon/dt \le 10^{19}~ erg~ g^{-1} s^{-1}$.}}
%%%%%%%%%%%%%%%%%%%%%%%%%%%%%%%%%%%%%%%%%%%%%%%%%%%%%%%%%%%%%%%%%%%%

\section{{Conclusions}}

In summary, we found that the emission of KK graviton by plasmon-plasmon annihilation from SN 1987A puts the conservative bound on the effective scale $M_D$  of the large extra dimensional model in the case of $d=2$ and $3$. Taking a conservative estimate of the supernovae temperature $T=30$ MeV and plasmon mass $m_A=19$ MeV (equal to the core plasma frequency $\om_o$), we find 
$M_D > 22.9$ TeV for $d=2$ and $M_D > 1.38$ TeV for $d=3$. No bound on $M_D$ follows from the present analysis for $d \ge 4$. 

%%%%%%%%%%%%%%%%%%%%%%%%%%%%%%%%%%%%%%%%%%%%%%%%%%%%%%%%%%%%%%%%%%%%%%%%%%%%
\section{{Acknowledgement}}
The authors are grateful to Professor Ramesh Kaul of Institute of Mathematical Sciences, Chennai for useful discussions.
%%%%%%%%%%%%%%%%%%%%%%%%%%%%%%%%%%%%%%%%%%%%%%%%%%%%%%%%%%%%%%%%%%%%%%%%%%%%
\appendix
\section{Several terms in Eq.~\ref{Eq:cross}}
\vspace*{-0.25in}
\bea
T_1 &=& \frac{1}{12 m_A^4} \left( S^4 - \frac{S^5}{m_{\vec {n}}^2} +\frac{S^6}{m_{\vec {n}}^4}  \right). \nonumber \\
T_2 &=& \frac{1}{6 m_A^2} \left( 5 \frac{S^4}{m_{\vec {n}}^2} - 
4 \frac{S^5}{m_{\vec {n}}^4}  \right). \nonumber \\
T_3 &=& m_A^2 \left( 12 S - \frac{20}{3} \frac{S^2}{m_{\vec {n}}^2} \right). 
\nonumber \\
T_4 &=& 16 m_A^4. \nonumber \\
T_5 &=& \frac{1}{3} \left( 14 S^2 - \frac{S^3}{m_{\vec {n}}^2} +4 \frac{S^4}{m_{\vec {n}}^4}  \right). \nonumber 
\eea

%\newpage

\end{document}